\documentclass[multphys,vecphys]{svmult}

\usepackage{makeidx}         
\usepackage{graphicx}        
\usepackage{multicol}        
\usepackage[bottom]{footmisc}

\makeindex             

\begin{document}

\title*{A new chemodynamical tool to study the evolution of galaxies in the
  local Universe: a quick and accurate numerical technique to compute
  gas cooling rate for any chemical composition}
\titlerunning{A new chemodynamical tool\dots}
\author{Nicolas Champavert\inst{1}\and
Herv\'e Wozniak\inst{1}}
\institute{Universit\'e Lyon 1, CRAL, Observatoire de Lyon, 9 avenue Charles Andr\'e, Saint-Genis Laval cedex, F-69561, France
\texttt{champavert@obs.univ-lyon1.fr}}
\maketitle

\begin{abstract}
We have developed a quick and accurate numerical tool to compute gas cooling whichever
its chemical composition. 
\end{abstract}

\section{Introduction}
\label{sec:1}
Metal abundances in galaxies vary from place to place and with
time. These variations reflect the star formation history within galaxies 
because the chemical enrichment results from synthesis of heavy elements by
successive generations of stars.
Gas radiative cooling is very sensitive to the ISM chemical composition.
Cooling influences both the gas dynamics and the star formation rate.
Therefore we need to follow self-consistently dynamical and chemical evolution.
We have developed a method to compute in a very short CPU time accurate
cooling functions dependent on the exact abundances of elements.
It will be used in high resolution chemodynamical simulations of galaxies.

\section{Description of the method}
\label{sec:2}
\begin{figure}[t]
\centering
\includegraphics[width=5.8cm]{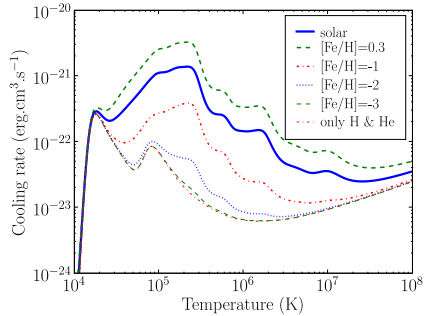}
\caption{Cooling rates as a function of temperature for different
  metallicities computed with Mappings~III. The abundances used in
  calculations are taken from \cite{s&d93}}
\label{cooling_comparisons}
\end{figure}

It is well known that cooling rates depend on gas chemical composition.
A gas with solar composition has a cooling rate greater than without metals
by more than an order of magnitude for some temperatures
(Fig.~\ref{cooling_comparisons}).
Cooling rates are generally interpolated between rates computed for different
metallicities with solar abundance ratios (see \cite{b&h89} for instance) or
with solar abundance ratios and enhanced abundances for some elements (see
\cite{s&d93} for instance).
The gas abundances are obviously not always solar because
the gas ejected from stars by winds and SNe has a non-solar composition.
Hence we need to take into account the real abundances of each element in
the gas to build more realistic cooling functions whichever the chemical
composition. However, computing cooling rates with Mappings~III for each gas
particle at each timestep of a chemodynamical simulation is a very CPU time
consuming task. We thus have developped a recipe to reconstruct the cooling
curve on the fly.

We consider the ISM as an optically thin gas in collisional
ionization equilibrium.
Calculations are performed in the temperature range $10^4$\,K to
$10^8$\,K using Mappings~III, successor of Mappings~II whose cooling
computations details are described in \cite{s&d93}.
All the abundances of elements are relative to hydrogen. The hydrogen density
is 1.0\,cm$^{-3}$, independent of temperature. 
We drop the temperature dependence of cooling rates ($\Lambda(T)$) for the
sake of legibility.

Many cooling processes depend on the density of electrons and on the
abundances of elements. A simple hypothesis is to suppose that cooling
rates are proportional to these two quantities. Thus, we can try to reconstruct the
total cooling curve $\Lambda_{tot}$ with a linear combination of individual curves for
each element.
We first need to compute these individual curves.
We obtain $\Lambda_H$ with a purely hydrogen gas.
For all the heavier elements used for calculations in Mappings~III (He, C, N,
O, Ne, Na, Mg, Al, Si, S, Cl, Ar, Ca, Fe, Ni), we have calculated cooling
rates for mixtures of hydrogen and one element $X_i$ at its solar abundance
$\Lambda_{X_i,H}$.
Individual cooling curves $\Lambda_{X_i}$ have then been obtained
substracting $\Lambda_H$ from each $\Lambda_{H,X_i}$: $\Lambda_{X_i} =
\Lambda_{H,X_i} - \Lambda_H$.
For a given chemical composition, we then compute the quantity $\alpha_{X_i}$
for all the elements previously listed, where
$\alpha_{X_i}$ is the abundance of element $X_i$ 
normalized by its solar abundance: $ \alpha_{X_i} = n_{X_i}/n_{X_{i\odot}}$.
For helium, our hypothesis is too simple. Its cooling rate cannot be
considered proportional to its abundance with a good approximation as there
are too strong non-linearities.
We choose to linearly interpolate in a grid of precomputed curves for
different abundances
to obtain the helium cooling curve $\Lambda_{He,\alpha_{He}}$.
As hydrogen and helium are the two most abundant elements, we assume that
they are the only sources of electrons. We indeed consider that the contribution of
the metals are negligible because of their much lower abundances in the ISM.
By construction, our individual cooling functions for metals take into account
the electrons of the element and hydrogen. Thus we only need to add the
contribution of electrons provided by the ionization of helium
$ne\left(He\right)$.
We add the terms for hydrogen, helium and metals to build $\Lambda_{tot}$ so
that we finally obtain the following approximation:
$  \Lambda_{tot} \approx \Lambda_H + \Lambda_{He,\alpha_{He}} +
  \left(1+ne\left(He\right)\right)\sum_{metals}\alpha_{X_i}\Lambda_{X_i}$

\begin{figure}[t]
\centering
\includegraphics[width=5.8cm]{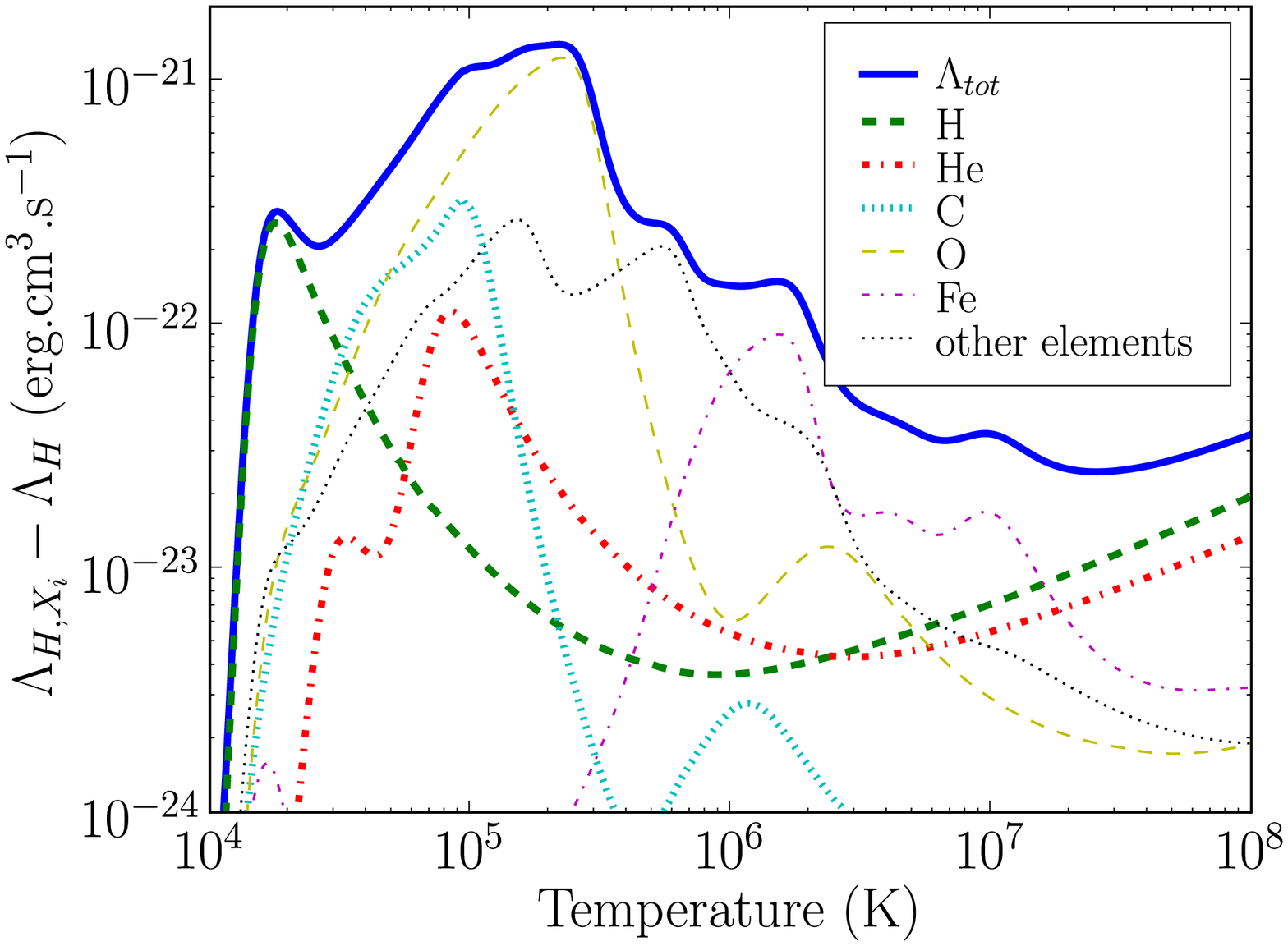} 
\includegraphics[width=5.8cm]{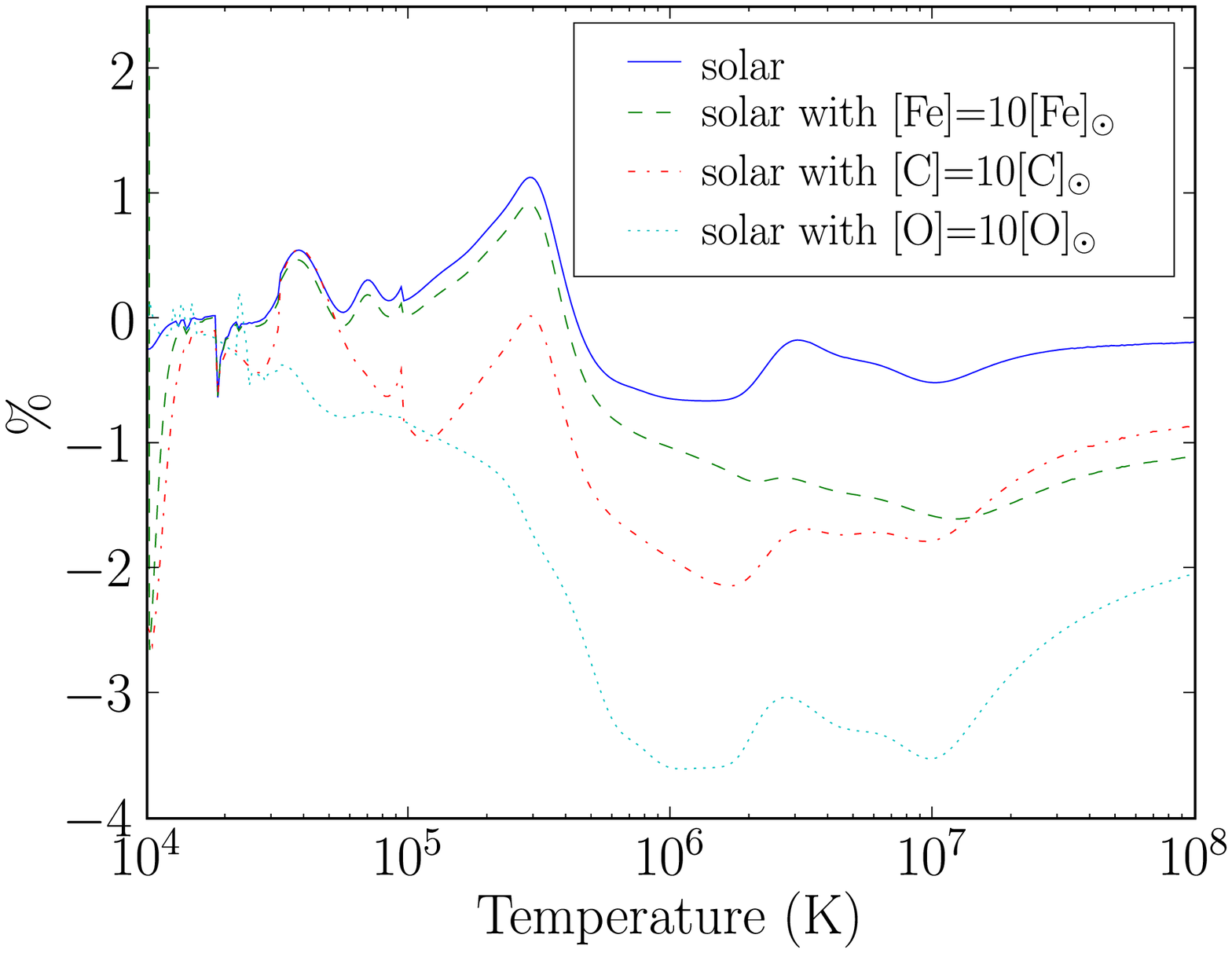} 
\caption{(\textbf{left panel}) Cooling rates for several elements at solar
  abundances. The curve for other elements is the sum of the contributions of
  N, Ne, Na, Mg, Al, Si, S, Cl, Ar, Ca and Ni. The solid broad line $\Lambda_{tot}$ is
  the cooling curve for solar abundances reconstructed using our recipe.
(\textbf{right panel}) Percentage
  of relative error made when reconstructing cooling functions with our recipe for
  different compositions
}
\label{cooling_elements}
\end{figure}

To check the accuracy of our tool, we have compared cooling curves computed
with Mappings~III and the ones reconstructed with our recipe for various
chemical compositions.
We choose a solar composition to test a well-known chemical composition.
Carbon, oxygen and iron are three of the most important coolants
(Fig.~\ref{cooling_elements}, left panel). Furthermore, they are three of the
most abundant elements in stellar winds and supernovae ejecta.
Thus, we increase by a factor of 10 the abundance of these three important
coolants. This factor is somewhat arbitrary and was only chosen to
test the accuracy of our recipe in the case of a very high-enhanced gas.
In all cases, the relative errors remain below a few percent
(Fig.~\ref{cooling_elements}, right panel).
The errors due to the computation of cooling rates are thus comparable to the
others coming from, for instance, the hydrodynamical scheme.

\printindex
\end{document}